\documentstyle[12pt,epsfig]{article}

\begin{document}   

\def\beq{\begin{equation}}
\def\eeq{\end{equation}}
\def\eq{\beq\eeq}
\def\beqn{\begin{eqnarray}}
\def\eeqn{\end{eqnarray}}
\relax
\def\ba{\begin{array}}
\def\ea{\end{array}}
\def\squ{\tilde{Q}}
\def\tb{\mbox{tg$\beta$}}
\def\hH{{\cal H}}
\def\ra{\rightarrow} 
\newcommand{\lsim}{\raisebox{-0.13cm}{~\shortstack{$<$ \\[-0.07cm] $\sim$}}~}
\newcommand{\gsim}{\raisebox{-0.13cm}{~\shortstack{$>$ \\[-0.07cm] $\sim$}}~}
\newcommand{\s}{\\ \vspace*{-2mm} }

\begin{flushright}
PM/97--44 \\
GDR--S--003
\end{flushright}

\vspace{1cm}

\begin{center}

{\large\sc {\bf Higgs Boson Production in Association with Scalar Top Quarks}}

\vspace*{3mm}

{\large\sc {\bf at Proton Colliders}}

 \vspace{1cm}

{\sc A.~Djouadi, J.-L. Kneur and G.~Moultaka}

\vspace{1cm}

Physique Math\'ematique et Th\'eorique, UMR--CNRS, \\
Universit\'e Montpellier II, F--34095 Montpellier Cedex 5, France.

\end{center}

\vspace*{2cm}

\begin{abstract}

\noindent 
We investigate the production of Higgs particles in association with  the
supersymmetric scalar partners of the top quark at proton colliders. In the 
minimal supersymmetric extension of the Standard Model, the cross sections for
the production of the lightest neutral Higgs boson in  association with top
squark pairs can be rather large, substantially  exceeding the rate for the
associated Higgs boson production with top quarks. If the lightest top squark
is not much heavier than the top quark, this process  will enhance the
potential of the CERN Large Hadron Collider to discover the lightest neutral
Higgs boson, and will open a window to the  study of the Higgs--stop coupling,
the potentially largest electroweak  coupling in the supersymmetric theory.  

\end{abstract}

\newpage

Supersymmetric theories (SUSY)~\cite{R1} are the best motivated extensions of
the Standard Model (SM) of the electroweak and strong interactions. They
provide an elegant way to break the electroweak symmetry and to stabilize the
huge hierarchy between the Grand Unification  and the Fermi scales. In the
Minimal Supersymmetric extension of the Standard Model (MSSM), the
SU(2)$\times$U(1) gauge symmetry is broken with two Higgs--doublet fields, 
leading to the existence of five physical states: two CP--even Higgs bosons 
$h$ and $H$, a CP--odd Higgs boson $A$ and two charged Higgs particles $H^\pm$
\cite{R2}. \s

In the theoretically well motivated models, such as Supergravity models
\cite{R3}, the MSSM Higgs sector is in the so called decoupling regime 
\cite{R4} for most of the SUSY parameter space allowed by present data 
constraints \cite{R5}: the heavy CP--even, the CP--odd  and the charged Higgs
bosons are rather heavy and almost degenerate in mass, while the lightest 
neutral CP--even Higgs  particle reaches its maximal allowed mass value 
$ M_h \lsim $ 60--130 GeV \cite{mh,lep2} depending on the SUSY parameters. 
In this scenario, the $h$ boson  has almost the same properties as the SM 
Higgs boson and would be the sole Higgs particle accessible at the next 
generation of colliders. \s

At the CERN Large Hadron Collider (LHC), the most promising channel \cite{R6} 
for detecting such a Higgs particle is the rare decay into two photons, $h
\rightarrow \gamma \gamma$, with the Higgs particle dominantly produced  via
the top quark loop mediated gluon--gluon fusion mechanism $gg  \rightarrow h$
\cite{R7}. The two LHC collaborations expect to detect the narrow $\gamma
\gamma$ peak in the entire intermediate Higgs mass range, 80 $\lsim M_h
\lsim 130$ GeV, with an integrated luminosity $\int {\cal L} \sim 100$ 
fb$^{-1}$ corresponding to one year LHC running \cite{R6}. \s

Two other channels can be used to detect the Higgs particle in this mass 
range:  the production in association with a $W$ boson, $pp \ra hW$ \cite{R8}, 
or  in association with top quark pairs, $pp \ra \bar{t}t h$ \cite{R9}, with 
the $t$ quarks decaying into $b$ quarks and $W$ bosons [for the latter process,
the Higgs boson detection with $h \ra b \bar{b}$ final states looks also
promising; see Ref.~\cite{bbh} for instance]. Although the cross sections are
smaller  compared to the $gg \ra h$ case, the background cross sections are
also  small if one requires a lepton from the decaying $W$ bosons as an
additional  tag, leading to a significant signal. Furthermore, the cross
section $\sigma( pp \ra \bar{t}th)$ is directly proportional to the top--Higgs
Yukawa coupling,  the largest electroweak coupling in the SM. This process
would therefore allow  the measurement of this parameter, and the experimental
test of a fundamental  prediction of the SM: the masses of fermions and gauge
bosons are generated through  the Higgs mechanism.  \s

In this paper, we point out that  in supersymmetric theories, which predict the
existence of scalar partners to  each SM chiral fermion, an additional process
might provide a new important  source for Higgs particles: the associated
production with the scalar partners  of the top quark, 
\begin{eqnarray}
pp \ra gg + q \bar{q} \ra \tilde{t} \tilde{t}h
\end{eqnarray}
The reason is twofold: \s

$(i)$ The current eigenstates, $\tilde{t}_L$ and $\tilde{t}_R$, mix to give 
the mass eigenstates $\tilde{t}_1$ and $\tilde{t}_2$ \cite{Ellis} which are 
obtained by diagonalizing the following mass matrix
\begin{equation}
{\cal M}^2_{\tilde{t}} = \left(
  \begin{array}{cc} m_{\tilde{t}_L}^2 + m_t^2+D_L  & m_t \, \tilde{A}_t \\
                    m_t \tilde{A}_t & m_{\tilde{t}_R}^2 + m_t^2+D_R
\end{array} \right)
\end{equation}
where in the off--diagonal entries $\tilde{A}_t= A_t - \mu/\tb$, with  $\tb$
the ratio of the vacuum expectation values of the two--Higgs fields  which
break the electroweak symmetry, and $A_t$ and $\mu$ the soft--SUSY  breaking
trilinear stop coupling and Higgs mass parameter, respectively. 
$m_{\tilde{t}_L}$ and
$m_{\tilde{t}_R}$ are the left-- and right--handed soft--SUSY breaking top 
squark masses  which, in models with universal scalar masses at the GUT scale, 
are approximately equal to the common squark mass $m_{\tilde{q}}$; the
$D$--terms in units of $M_Z^2 \cos 2\beta$ are: $D_L=\frac{1}{2}-\frac{2}{3}
\sin^2\theta_W$ and $D_R=\frac{2}{3} \sin^2\theta_W$. The mixing angle
$\theta_{\tilde{t}}$ is proportional to $m_t \tilde{A}_t$ and can be very
large, leading to a scalar top quark $\tilde{t}_1$ much  lighter than the
$t$--quark and all other scalar quarks.
The reaction $pp \ra \tilde{t}_1 \tilde{t}_1 h$ can be, therefore, more 
phase--space favored than the corresponding SM--like process $pp  \ra 
\bar{t}th$. \s

$(ii)$ Normalized to $2(\sqrt{2}G_F)^{1/2}$, the couplings of the top squarks 
to the lightest Higgs boson read in the decoupling regime, 
\begin{eqnarray}
g_{h \tilde{t}_1 \tilde{t}_1 } &=& - \frac{1}{2} \cos 2\beta \left[ 
\cos^2 \theta_{\tilde{t}} - \frac{4}{3} \sin^2 \theta_W \cos 2
\theta_{\tilde{t}} \right] \nonumber \\
&& - \frac{m_t^2}{M_Z^2} - \frac{1}{2} \sin 2\theta_{\tilde{t}} 
\frac{m_t \tilde{A}_t } {M_Z^2}  \nonumber \\
g_{h \tilde{t}_2 \tilde{t}_2 } &=& - \frac{1}{2} \cos 2\beta \left[ 
\sin^2 \theta_{\tilde{t}} + \frac{4}{3} \sin^2 \theta_W \cos
2\theta_{\tilde{t}}  \right] \nonumber \\
&& - \frac{m_t^2}{M_Z^2} + \frac{1}{2} \sin 2\theta_{\tilde{t}} 
\frac{m_t \tilde{A}_t } {M_Z^2}
\end{eqnarray}
involving 
components which are proportional to $\tilde{A}_t$. For large values of the
parameter $\tilde{A}_t$ [which incidentally make the $\tilde{t}$ mixing 
angle  maximal $|\sin 2 \theta_{\tilde{t}}| \simeq 1$], the latter terms can
strongly  enhance the Higgs couplings to top squarks, and make it larger than
the  top quark coupling of the Higgs boson, $g_{htt} \propto m_t/M_Z$. Both
components   would result in an enhancement of the $pp \ra \tilde{t} \tilde{t}
h$ rate  compared to the $pp \ra \bar{t}th$ cross section. \s

In this letter, we analyze the associated production of Higgs bosons with 
pairs of top squarks at the LHC with a center of mass energy $\sqrt{s}=14$ 
TeV. We will concentrate on the case of the lightest $h$ boson 
of the MSSM in the decoupling regime, and  discuss only the production in
association with light top squarks. A more detailed discussion will be 
postponed to Ref.~\cite{later}. 
\vspace*{2mm}

\noindent
\begin{figure}[htb]
\begin{center}
\mbox{ 
\psfig{figure=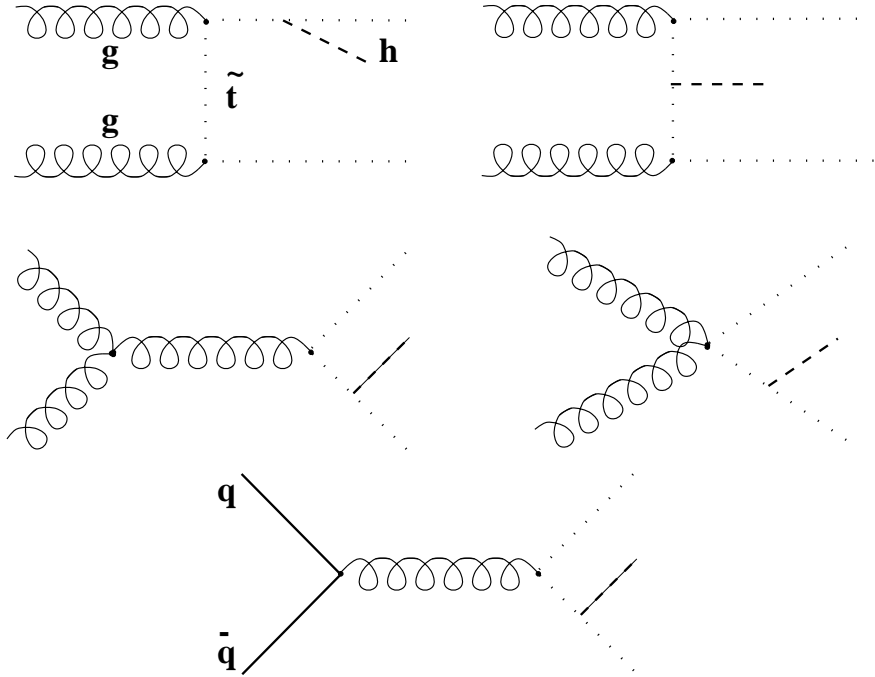,width=8cm,angle=-90}}
\end{center}
\caption[]{Generic Feynman diagrams for the production of the $h$ boson 
in association with top squarks via $gg$ fusion and $q\bar{q}$ 
annihilation.}
\end{figure}

At lowest order, i.e. at ${\cal O}( G_F \alpha_s^2)$, the process is  initiated
by the Feynman diagrams shown in Fig.~1. There are 10 diagrams  for the $gg$
fusion mechanism [including those  with the quartic gluon--squark interaction
and the three--gluon vertex] once the various  possibilities for emitting the
Higgs boson from the squark lines and the  crossing of the two gluons are
added, and 2 diagrams for the $q\bar{q}$  annihilation process. The ${\cal
O}(G_F^3)$ contribution from $\gamma$ and  $Z$--boson exchange diagrams, 
as well as
the flavor changing contribution with a  gluino exchange diagram are negligible
since they are suppressed, respectively, by the  additional weak coupling 
factor and the very small mixing between  light quarks and top squarks through 
the gluino interaction, the mixing  being due to weak interactions again. 
Due to the
larger gluon luminosity at  high energies, the contribution of the $gg$--fusion
diagrams is much larger  than the contribution of the $q\bar{q}$ annihilation
diagrams at LHC  energies, the difference  being almost two orders of magnitude
for relatively small $\tilde{t}$ mass  values. The analytical expressions of
the partonic cross section are rather lengthy, and will be given elsewhere 
\cite{later}. \s

To simplify our numerical analysis, we will assume the left--  and right--handed
stop mass parameters to be equal, $m_{\tilde{t}_L}= m_{\tilde{t}_R} \equiv
m_{\tilde{q}}$ as is approximately the case in GUT  scenarios. For
illustration, we have chosen the values $\tb=2$ and $30$.  The $h$ boson mass
is then calculated  as a function of $m_{\tilde{q}}, A_t$ and $\mu$ [$M_h$ 
is only marginally affected by the variation of the latter parameter]  
with the pseudoscalar Higgs boson mass fixed to $M_A=1$ TeV, and with 
the full radiative corrections in the improved effective  potential 
approach \cite{lep2} included. The top quark mass is fixed to $m_t=175$  
GeV, and the most recent CTEQ4 parameterizations of the structure 
functions \cite{CTEQ} are chosen. \s

In Fig.~2, the $pp \ra \tilde{t}_1 \tilde{t}_1h$ cross section [in pb] is 
displayed as a function of the lightest $\tilde{t}$ mass for the value 
$\tb=2$, in the case of no--mixing $\tilde{A}_t =0$ [$A_t=200, \mu=400$ GeV],
moderate mixing [$A_t=500$ and $\mu=100$ GeV] and large mixing 
[$A_t=1.5$ TeV and $\mu=100$ GeV]. Note for comparison, that  
the cross section for the standard--like $pp \ra \bar{t} t h$ process 
is of the order of 0.6 pb for a Higgs boson mass $M_h \simeq 100$ GeV 
\cite{tth}. \s

In the case where there is no mixing in the stop sector, $\tilde{t}_1$  and
$\tilde{t}_2$ have almost the same mass 
[which, up to the small contribution of the
D--terms, is constrained to be larger than $m_t^{\rm \overline{MS}}$] and
approximately the same couplings to the Higgs boson since the $m_t^2/M_Z^2$
components are dominant. The cross  section in Fig.~2,  which should be then
multiplied by a factor of two to take  into account the production of both
squarks, is comparable to the $pp \ra t\bar{t} h$ cross section in the  low
mass range $m_{\tilde{t}}\lsim  200$ GeV. In scenarios where the  $\tilde{t}$
masses are related to the masses of the light quark partners,  $m_{\tilde{q}}$,
the mass range for which the cross section is rather large  is, however, ruled
out by present experimental constraints on $m_{\tilde{q}}$ \cite{R5}.  \s

For intermediate values of $\tilde{A}_t$ the two components of the  $h
\tilde{t}_1 \tilde{t}_1$ coupling interfere destructively and partly 
cancel each other, resulting in a rather small cross section, unless
$m_{\tilde{t}_1} \sim {\cal O}(100)$ GeV. For some value of $\tilde{A}_t$
the $h\tilde{t}_1 \tilde{t}_1$ coupling is zero and the cross section 
vanishes. \s

In the large mixing case, $\tilde{A}_t \sim 1.5$ TeV, 
$\sigma(pp \ra \tilde{t}_1 
\tilde{t}_1 h)$ can be very large. It is above the rate for the standard 
process $pp \ra \bar{t}th$ for values of $m_{\tilde{t}_1}$ smaller than 220 GeV.
If  $\tilde{t}_1$ is lighter than the top quark, the $\tilde{t}_1 \tilde{t}_1 
h$ cross section significantly exceeds the one for $\bar{t}th$ final states.
For instance, for $m_{\tilde{t}_1}=140$ GeV, corresponding to $M_h\sim 76$ GeV, 
$\sigma( pp \ra \tilde{t}_1
\tilde{t}_1 h)$ is an order of magnitude larger than $\sigma(pp \ra t\bar{t}h)$.
Note that large values of $\tilde{A}_t$,  for which in GUT scenarios one needs
a sizeable common squark mass parameter  $m_{\tilde{q}}$  to avoid color
breaking minima, correspond to the ``maximal--mixing" scenarios which maximize
the $h$ boson mass \cite{lep2}. \s

\begin{figure}[htb]
\begin{center}
\mbox{
\psfig{figure=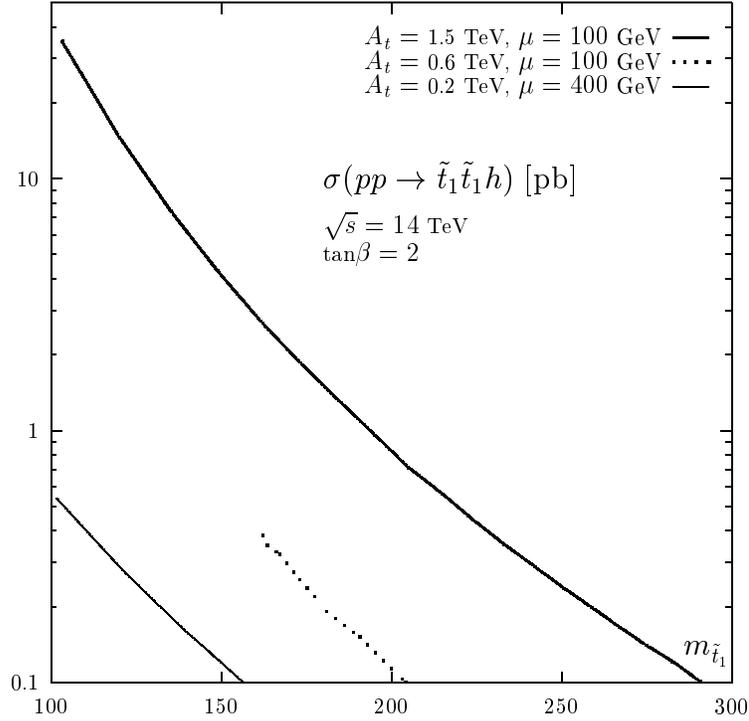,width=14cm}}
\end{center}
\vspace*{-8cm}
\caption[]{The production cross section $\sigma (pp \ra \tilde{t}_1 
\tilde{t}_1 h$) [in pb] as a function of the $\tilde{t}_1$ mass and three 
sets of $A_t$ and $\mu$ values; $\tb$ is fixed to $\tb=2$.} 
\end{figure}

In Fig.~3, we fix the lightest top squark mass to $m_{\tilde{t}_1} =165$ GeV 
$ \sim m_{t}^{\rm \overline{MS}}$ 
and display the $pp \ra gg+q\bar{q} \ra \tilde{t}_1 \tilde{t}_1h$ cross 
section as a function of $\tilde{A}_t$ for two values of $\tb=2$ 
and $\tb=30$. For comparison, the $*$ and $\bullet$ give the standard--like 
$pp \ra \bar{t}th$ cross section for $M_h=100$ GeV and $\tb=2$ and 30, 
respectively. For $\tb=30$ the cross
section is somewhat smaller than for $\tb=2$, a mere consequence of the
increase of the $h$ boson mass with $\tb$ \cite{lep2}.  As can be seen again,
the production cross  section is substantial for the no--mixing case, rather
small for intermediate mixing [becoming negligible for $\tilde{A}_t$
values between 200 and 400 GeV], and then becomes very large, exceeding the 
reference cross section for values of $\tilde{A}_t$  above $\sim 1$ TeV.  \s

\begin{figure}[htb]
\begin{center}
\mbox{
\psfig{figure=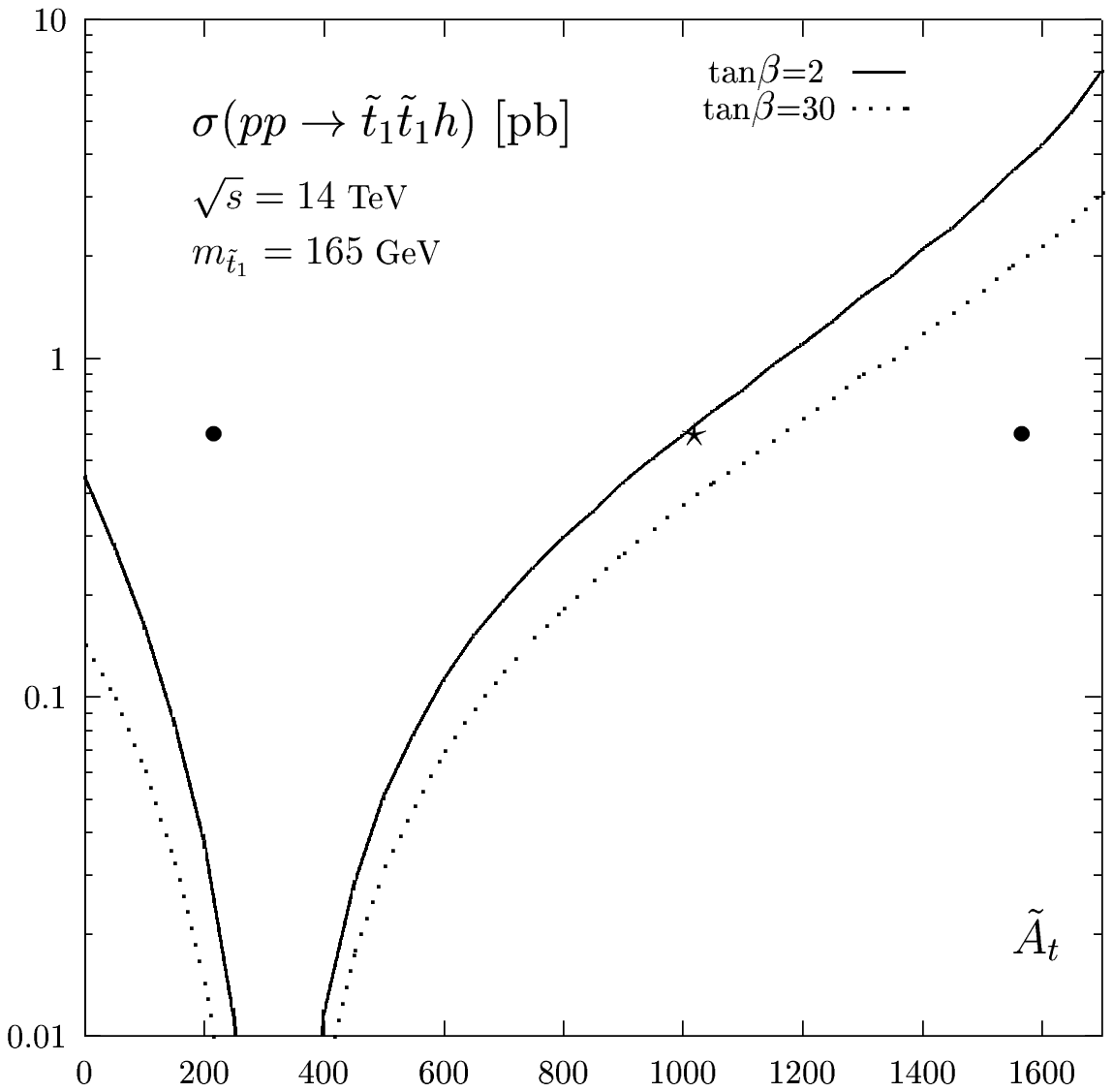,width=14cm}}
\vspace*{-8cm}
\end{center}
\caption[]{The cross section $\sigma(pp \ra \tilde{t}_1\tilde{t}_1h
$) [in pb] as a function of $\tilde{A}_t$ for fixed $m_{\tilde{t}_1}=165$ GeV
and for $\tb=2,30$. $*(\bullet)$ is for $\sigma(pp \ra t\bar{t}h$) with 
$M_h=100$ GeV and $\tb=2(30)$.} 
\end{figure}

Note that for fixed $\tilde{t}$ mass and coupling, the cross section becomes
smaller for larger values of $\tilde{A}_t$, if $\tilde{A}_t \lsim \sqrt{6} 
m_{\tilde{q}}$, because $M_h$ increases \cite{lep2} and the process is less 
favored by phase--space; in the reverse situation, $\tilde{A}_t \gsim \sqrt{6} 
m_{\tilde{q}}$, the $h$ boson mass will start decreasing with increasing 
$\tilde{A}_t$ [reaching values below $M_h \lsim 60$ GeV when $\tilde{A}_t 
\sim 1.75$ and 2 TeV, for $\tb=2$ and 30 respectively] and the phase--space 
is more favorable to the reaction.  

\smallskip


Let us now discuss the signal for the $pp \ra \tilde{t}_1 \tilde{t}_1 h$
process. In most of the parameter space, the top squark will decay into a $b$
quark and a chargino, $\tilde{t}_1 \ra b\chi^+$, if  $m_{\tilde{t}_1} <
m_t+m_{\chi_1^0}$  where $\chi_1^0$ is the lightest  SUSY particle (LSP), or
into a $t$ quark and the LSP, $t \ra t \chi_1^0$, in the opposite case
\cite{stdecay}. In the interesting region where  the cross section $\sigma(pp 
\ra \tilde{t}_1  \tilde{t}_1 h)$ is large, i.e. for relatively light 
$\tilde{t}_1$, the decay  mode  $\tilde{t}_1 \ra b \chi^+$ is dominant, unless 
the mass difference  $m_{\tilde{t}_1} -m_{\chi_1^+}$ is very small, in which 
case the loop induced  decay, $\tilde{t}_1 \ra c \chi^0_1$, can become 
competitive.  
In this region, the strong decay into gluinos does not occur. Assuming that 
the  partners of the leptons are heavier than
the lightest chargino, $\chi_1^+$  will mainly decay into the LSP and a real or
virtual $W$ boson, leading to the  decay
\begin{eqnarray}
\tilde{t}_1 \ra bW^+ \ + \ {\rm missing \ energy}
\end{eqnarray}
This is the same topology as in the case of the top quark decay, $t \ra 
bW^+$, except that in the case of the top squark there is a large amount of 
missing energy due to the undetected LSP. If sleptons are also relatively 
light, charginos decays will also lead to $l \nu \chi_1^0$ final states. 
The only difference between the final states generated by the $\tilde{t}
\tilde{t}h$ and $t\bar{t}h$ processes, will be due to the softer energy 
spectrum of the charged leptons coming from the chargino decay in the former
case, because of the energy carried by the invisible LSP. \s

The Higgs boson can be tagged through its $h \ra \gamma \gamma$ decay mode.
In the decoupling limit, and for light top squarks and large $\tilde{A}_t$
values, the branching ratio for this mode can be substantially enhanced
compared to the SM Higgs boson \cite{brh}, because of the additional
contributions of the $\tilde{t}$--loops which interfere constructively with 
the dominant $W$--loop contribution. Therefore, $\gamma \gamma$+ charged 
lepton events can be much more copious than in the SM, and the contributions 
of the $pp \ra \tilde{t} \tilde{t} h$ process to these events can render the 
detection of the $h$ boson much easier than with the process $pp \ra t 
\bar{t}h$ alone. \s

Although a  detailed Monte-Carlo analysis, which is  beyond the scope of this
letter, will be required to  assess the importance of this signal and to
optimize the cuts needed not to dilute the contribution of the $\tilde{t}
\tilde{t}h$ final states, it is clear that in a substantial area of the
MSSM parameter space, the contribution of the top squark to the $\gamma \gamma
l^\pm$ signal can significantly enhance the potential of the LHC to discover
the lightest MSSM Higgs boson in this channel. Note that an excess of $\gamma 
\gamma l^\pm$ events  would indicate the presence of additional contributions 
from top squarks to the Higgs boson production. This would be a new and very 
interesting means to search for top squarks at the LHC, which due to the large 
QCD background from $\bar{t} t$ production, are otherwise difficult to detect 
in other channels. Last but not least and as a welcome bonus, an excess of 
$\gamma \gamma l^\pm$  
events, compared to the expected rate from the $pp \ra t\bar{t}h$ process, 
will allow to measure the $h\tilde{t} \tilde{t}$ coupling, the largest 
electroweak coupling in the MSSM, opening thus a window to probe directly the 
soft--SUSY breaking scalar potential.  \s

In summary, we have calculated the cross section for the production of the
lightest CP--even neutral MSSM Higgs particle in association with  the scalar
partners of the top quark at proton colliders, $pp \ra gg + q \bar{q} \ra
\tilde{t} \tilde{t}h$.  The cross section can  substantially exceed the rate
for the associated  production with top quarks, especially for large values of
the off--diagonal entry of the $\tilde{t}$  mass matrix which, at the same
time, makes $\tilde{t}_1$ much lighter than the  other squarks and increases
its coupling to the $h$ boson.  This process can  strongly enhance the
potential of the LHC to discover the $h$ boson in the  $\gamma \gamma+l^\pm$
channel, and would open a window for the determination  of the important
$\tilde{t} \tilde{t} h$ coupling. Finally, this reaction  could be a new
channel to search for relatively light top squarks at hadron  colliders. 

\bigskip

\noindent {\bf Acknowledgments:} \\
This work is partially supported by the French GDR--Supersym\'etrie.

\end{document}